\documentclass[aps,prl,twocolumn,superscriptaddress,footnotebib]{revtex4-1}
\usepackage{amsmath}
\usepackage{amssymb}
\usepackage{amsfonts}
\usepackage{bm}
\usepackage{color}
\usepackage{graphicx}
\usepackage{epstopdf}
\usepackage{epsfig}
\usepackage[naturalnames]{hyperref}
\usepackage{soul}
\usepackage{hypcap}
\usepackage{verbatim}
\usepackage{tabularx}
\usepackage{bbm}
\usepackage{esvect}
\usepackage{subfigure}

\def\bea{\begin{eqnarray}}
\def\eea{\end{eqnarray}}
\def\nn{\nonumber}
\def\ba{\begin{array}}
\def\ea{\end{array}}
\def\Tr{\text{Tr}}
\def\nn{\nonumber}
\def\sgn{\text{sgn}}

\def\B{\textcolor{blue}}

\hypersetup{colorlinks=true, citecolor=blue, urlcolor=blue, linkcolor=blue}
\begin{document}
	
\title{Quantum error as an emergent magnetic field}
	
	\author{Shao-Kai Jian}
	\affiliation{Condensed Matter Theory Center and Joint Quantum Institute,
Department of Physics, University of Maryland, College Park, MD 20742, USA}

	\author{Chunxiao Liu}
	\thanks{chunxiaoliu@ucsb.edu}
	\affiliation{Department of Physics, University of California Santa Barbara, Santa Barbara, CA 93106, USA}

	\author{Xiao Chen}
	\thanks{chenaad@bc.edu}
    \affiliation{Department of Physics, Boston College, Chestnut Hill, MA 02467, USA}
    
    \author{Brian Swingle}
	\thanks{bswingle@umd.edu}
    \affiliation{Department of Physics, Brandeis University, Waltham, Massachusetts 02453, USA}
    \affiliation{Condensed Matter Theory Center and Joint Quantum Institute,
Department of Physics, University of Maryland, College Park, MD 20742, USA}
    
    \author{Pengfei Zhang}
	\thanks{PengfeiZhang.physics@gmail.com}
	\affiliation{Institute for Quantum Information and Matter and Walter Burke Institute for Theoretical Physics, California Institute of Technology, Pasadena, CA 91125, USA}


\begin{abstract}
    We investigate the effect of quantum errors on a monitored Brownian Sachdev-Ye-Kitaev (SYK) model featuring a measurement-induced phase transition that can be understood as a symmetry-breaking transition of an effective $Z_4$ magnet in the replica space.
    The errors describe the loss of information about the measurement outcomes and are applied during the non-unitary evolution or at the end of the evolution.
    In the former case, we find that this error can be mapped to an emergent magnetic field in the $Z_4$ magnet, and as a consequence, the symmetry is explicitly broken independent of the measurement rate. 
    R\'enyi entropies computed by twisting boundary conditions now generate domain walls even in the would-be symmetric phase at a high measurement rate. The entropy is therefore volume-law irrespective of the measurement rate. 
    In the latter case, the error-induced magnetic field only exists near the boundary of the magnet. 
    Varying the magnetic field leads to a pinning transition of domain walls, corresponding to error threshold of the quantum code prepared by the non-unitary SYK dynamics.
\end{abstract}
\maketitle

\B{\it Introduction.---}Pure quantum states correspond to states of maximal knowledge of a quantum system, and various kinds of ideal quantum dynamics preserve this property, for example, unitary dynamics or projective measurements in which the complete measurement record is retained. Quantum errors occur when this idealized dynamics is interrupted, typically leading pure states to evolve into mixed states. Such errors are ubiquitous in nature and play a fundamental role across a variety of disciplines. For instance, in condensed-matter physics and quantum information physics, errors are inevitable because any realistic system in a lab is coupled to its environment, typically resulting in entangled states in which the system alone is no longer in a pure state.

In the emerging field of hybrid dynamics~\cite{li2018quantum, li2019measurement, skinner2019measurement,choi2020quantum, gullans2020dynamical,chan2019unitary,zabalo2020critical, gullans2020scalable, li2020conformal, fan2020self, iaconis2020measurement, sang2020measurement, lavasani2021measurement, ippoliti2021entanglement, chen2020emergent, alberton2020trajectory, jian2020criticality, tang2021quantum, buchhold2021effective, bao2021symmetry, zhang2021syk, jian2021syk}, it was recently discovered that the quantum trajectories resulting from chaotic unitary evolution and repeated measurements exhibit a phase transition between a volume-law and an area-law entangled phase~\cite{li2018quantum, li2019measurement, skinner2019measurement, choi2020quantum, gullans2020dynamical}. On one hand, 
failure to retain the complete measurement record for each quantum trajectory destroys this phase transition. In this case, the system is effectively coupled to an inaccessible environment, leading to a trivial thermal volume-law phase regardless of the coupling strength. On the other hand, strictly following the quantum trajectory, the hybrid circuit can be interpreted as preparing a quantum error correction code~\cite{choi2020quantum, gullans2020dynamical}. In the context of error correction, the environment appears as an error that can cause the code to fail above a threshold~\cite{nielsen2001quantum,dennis2002topological}.

In this paper, we develop a theoretical framework to understand the effects of quantum errors in non-unitary dynamics. 
Concretely, we use the monitored Brownian Sachdev-Ye-Kitaev (SYK) chain~\cite{jian2021syk, kitaev2015simple, sachdev1993gapless, maldacena2016remarks, liu2018quantum, gu2017spread, huang2019eigenstate, zhang2020subsystem, haldar2020renyi, zhang2020entanglement, chen2020replica, garcia2021replica, zhang2021universal}, and model quantum errors by throwing away individual measurement records with some probability. 
Without errors, the model exhibits a measurement-induced phase transition that can be understood as a symmetry-breaking transition of an effective $Z_4$ magnet in the replica space~\cite{jian2021syk}. The subsystem entanglement
entropy corresponds to the free energy of topological defects created by the twisted boundary conditions. 
In the symmetry-breaking phase at a low measurement rate, domains with different orientations are enforced by the twisted boundary condition and are separated by domain walls with finite line tension, giving rise to volume-law entanglement entropy.
In the symmetric phase at a high measurement rate, the boundary condition can only change the free energy locally which leads to area-law entanglement entropy~\cite{EE}. 
We introduce errors in the non-unitary quantum dynamics (during the encoding process) or in the steady state (after the code is prepared), and analyze their effect on the entanglement scaling. 
In the former case, we find that quantum errors can be mapped to an emergent magnetic field in the $Z_4$ magnet. In the presence of such errors, the symmetry is explicitly broken independent of the measurement rate, and consequently, the measurement-induced phase transition is absent. 
In the latter case, the error induces a magnetic field near the boundary which leads to a pinning transition of domain walls. We argue that this is closely related to the error threshold problem for the quantum code generated by the non-unitary SYK dynamics. 
Notice that the connection between the pinning transition and the quantum error threshold has also been discussed in Refs.~\onlinecite{li2020statistical,gullans2020quantum,li2021entanglement}.
We also numerically calculate the subsystem entropy to show the absence of measurement-induced phase transition in the former, and the existence of the pinning transition in the latter.

\B{\it Model and setup.---}
We consider the following Brownian SYK Hamiltonian of $L$ and $R$ chains~\cite{saad2018semiclassical, sunderhauf2019quantum, liu2020non, jian2020note, jian2021syk},
\bea \label{eq:LRhamiltonian}
	& H = \sum_{x;a=L,R}  \Big( \sum_{ij}  i J_{a,ij}^{x,x+1}(t) \psi_{x,a,i} \psi_{x+1,a,j} \nn \\
		&+ i^{q/2} \sum_{j_1<...<j_q} U_{a,j_1...j_q}^x(t) \psi_{x,a,j_1}... \psi_{x,a,j_q}  \Big),
\eea
where $\psi_{x,a,i}$, $i=1,...,N$, denotes $i$-th of $N$ Majorana fermion at site $x$ of the $a \!=\! L, R$ chains. 
$L$ is the number of sites and periodic boundary conditions are assumed. The couplings $J_{a,ij}^{x,x+1}$ ($U_{a,j_1,...,j_q}^x$) correspond to hopping (interaction) and are independent Gaussian variables with mean zero and variances
\bea\label{eq:random}
& \overline{J_{a,ij}^{x,x+1}(t_1) J_{a',ij}^{x',x'+1}(t_2)} = \frac{J}{2N} \delta(t_{12})  \delta_{aa'} \delta^{x,x'}, \\
& \overline{U_{a,j_1...j_q}^{x}(t_1) U_{a',j_1...j_q}^{x'}(t_2)} = \frac{2^{q-2}(q-1)! U}{N^{q-1}} \delta(t_{12})  \delta_{aa'} \delta^{x,x'}. \nn
\eea

The system is under continuous monitoring given by measurement operators for the $i$-th Majorana fermion at site $x$ with probability $p$,
\bea \label{eq:measurement}
& M_1 =  \kappa_+ - \kappa_- \left(\frac12 + i \psi_{x,L,i}\psi_{x,R,i} \right), \\
& M_2 = s \left(\psi_{x,L,i} + i \psi_{x,R,i} \right),
\eea
where $\kappa_\pm \!=\! ( \frac{1 \pm \sqrt{1-4s^2}}{2} )^{1/2}$, $0 \!<\! s \!<\! \frac12$ is the measurement strength, and $ M_1^\dag M_1 + M_2^\dag M_2 = \mathbb{I}$. 
To model the error, we assume that when a measurement is performed, there is a probability $\gamma$ to lose the measurement record.

We are interested in calculating the quasi-$n$ entropy of bipartite system $A \bar A$~\cite{napp2019efficient},
\bea \label{eq:entropy}
&	S_A^{(n)} = \frac1{1-n} \log \frac{\mathbb E \Tr(\rho_{A}^n)}{\mathbb E\Tr(\rho)^n} = \frac1{1-n} \log \frac{\mathbb E \Tr(\rho^{\otimes n} X_A)}{\mathbb E\Tr(\rho^{\otimes n})},
\eea
where $\rho$ is the unnormalized density matrix, $X_A$ denotes the cyclic permutation acting on the subsystem $A$ of the $n$ replicated Hilbert space, and $\mathbb E$ denotes average over the Brownian and continuous monitoring evolution.
We focus on the quasi-2 entropy, so we need four contours that are denoted as: $1,2$ ($3,4$) denote the first (second) replica, and $1,3$ ($2,4$) denote the forward (backward) evolution.
When the measurement is implemented with a probability $p$, during a time step $\delta t$ the effective action in the four contours reads~\cite{suppl}
\bea \label{eq:measurement}
    &   (1-p) \mathbb{I} + p \sum_{i,j,\mu} w_{ij}^{(\mu)} M_i \otimes M_i^\dag \otimes M_j \otimes M_j^\dag \nn \\
    & = \exp \delta t \Big( - \frac{\mu}2 \sum_\alpha i G_{x, LR}^{\alpha\alpha} - \frac{\gamma \mu}2 (M^{12} + M^{34})  \Big),
\eea
where in the first line $w_{ij}^{(1)} \!=\! (1-\gamma) \delta_{ij}$ ($w_{ij}^{(2)} = \gamma$) corresponds to keeping (losing) the record. In the second line, $M^{\alpha\beta} \!=\!  G_{LL}^{\alpha\beta} \!+\! G_{RR}^{\alpha\beta} \!-\! i  G_{LR}^{\alpha\beta} \!+\! i G_{RL}^{\alpha\beta}$, $\alpha, \beta \!=\! 1,...,4$ denote the four contours and $\mu = p s^2 /\delta t$ is the measurement rate.

Including the Brownian Hamiltonian~(\ref{eq:LRhamiltonian}) and the measurement~(\ref{eq:measurement}) and integrating out the Gaussian variables, the effective action is
\bea \label{eq:action}
& \!-\! \frac{I}N = \frac12 \Tr \log [ (-1)^{\alpha+1} \delta^{\alpha\beta} \delta_{ab} \partial_t \!-\! \Sigma_{ab,x}^{\alpha\beta} ] - \frac12 \int \Sigma_{ab,x}^{\alpha\beta} G_{ab,x}^{\alpha\beta} \nn \\
& 	\!+\! \int \delta(t \!-\!t') \Big[ \frac{(-1)^{\alpha + \beta + 1}}{4} \delta_{ab} [ J G_{ab,x}^{\alpha\beta} G_{ab,x+1}^{\alpha\beta} + \frac{U}{2q} (2G_{ab,x}^{\alpha\beta})^{q} ] \nn \\
 & - \frac{\mu}2 \sum_\alpha i G_{x, LR}^{\alpha\alpha} - \frac{\gamma \mu}2 ( M^{12} + M^{34} ) \Big].
\eea
where $\Sigma_{ab,x}^{\alpha\beta}(t,t')$ is the self energy introduced to enforce $ G_{ab,x}^{\alpha\beta}(t,t') = \frac1N \sum_j \psi_{x,a,j}^\alpha(t) \psi_{x,b,j}^\beta(t')$. The summations over $x$, $a,b$ and $\alpha, \beta$ are implicit. We focus on $q=4$.

\begin{figure}
	\centering
    \includegraphics[width=0.45 \textwidth]{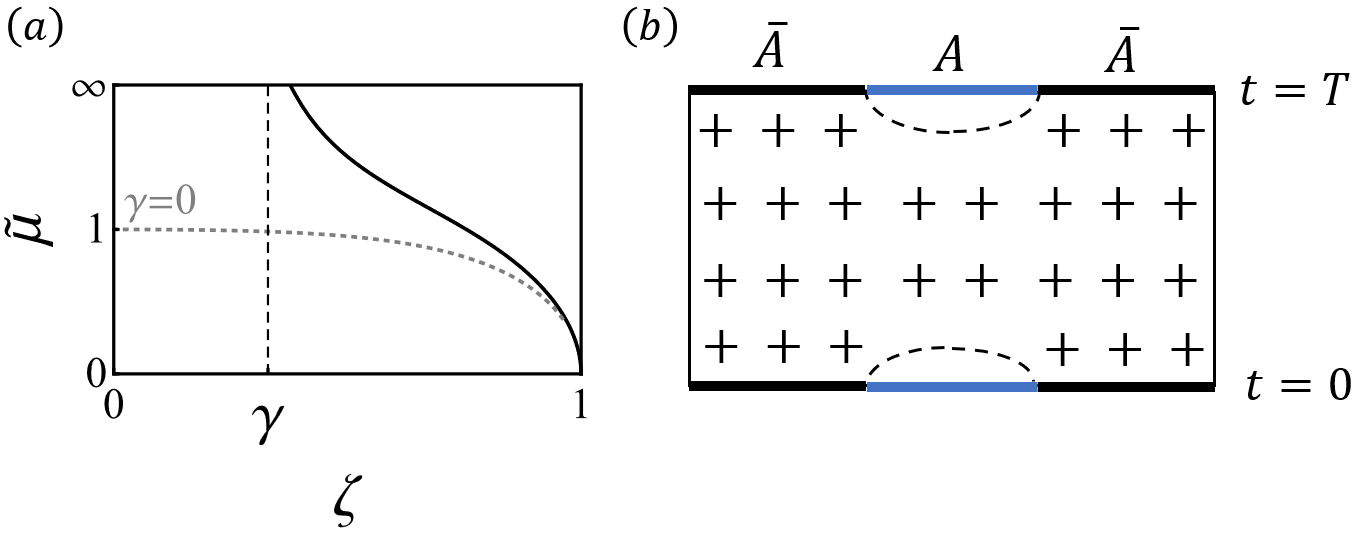}
	\caption{(a) An illustration of $\zeta$ determined by $\tilde \mu$ in (\ref{eq:zeta}). When $\gamma>0$, one can see that $\zeta > \gamma$.  
	The case for $\gamma =0$ is plotted by the dotted line for comparison. (b) A schematic plot of domain wall configurations. Throughout the paper, the $+$ and $-$ indicates $(\phi_1\!>\!0,\phi_2\!=\!0)$ and $(\phi_1\!=\!0,\phi_2\!>\!0)$, respectively, and the black (blue) thick line denotes the boundary pointing toward $+$ ($-$) at boundaries $t=0$ and $T$. The magnetic field points toward $+$ direction. \label{fig1}}
\end{figure} 

\B{\it Emergent magnetic field.---} The saddle-point solution to the large-$N$ action within a replica reads
\bea \label{eq:SYKq_saddle1}
    & \bar G = \frac{e^{- \frac{|t|}2 (J + U \zeta^{2} + \frac{\gamma\mu}\zeta)}}{2} [\sgn(t) \sigma^z - \zeta i \sigma^y - \frac{\mu \tau^y( 1 + \gamma \sigma^x)}{J + U \zeta^{2} + \mu \frac{\gamma}\zeta}],
\eea
where Pauli matrix $\sigma $ ($\tau$) acts on 1 and 2 contours ($L$ and $R$ chains).
The solution on $3,4$ contours is the same, consistent with the boundary condition without twist operators. 
The parameter $\zeta$ is given via the relation,
\bea \label{eq:zeta}
    & \tilde \mu = \frac{\zeta(1+ \tilde U \zeta^2)[ \gamma(1-\zeta^2) + \zeta \sqrt{(1-\zeta^2)(1-\gamma^2)}]}{\zeta^2 - \gamma^2},
\eea
with $\tilde U \!=\! \frac{U}J$ and $\tilde \mu \!=\! \frac\mu{J}$.

When $\gamma \!=\! 0$, (\ref{eq:action}) has $C_4 \!\times\! C_4$ symmetry~\cite{jian2021syk}, $G_{ab,x} \rightarrow O^{-1} G_{ab,x} O,$ where $O = e^{i(\theta_{13}\gamma_{13}+ \theta_{13} \theta_{24})}$ acts identically on the left and right chains. $\gamma_{ij}^{\alpha\beta} \!=\!  \delta_i^\alpha \delta_j^\beta -\delta_j^\alpha \delta_i^\beta$, and $\theta_{13}, \theta_{24} = \frac{n\pi}2$ ($n$ is an integer).
The relative rotation symmetry generated by $\gamma_{13} - \gamma_{24}$ is spontaneously broken by nonzero $\zeta$ in solution (\ref{eq:SYKq_saddle1}).
If no error occurs, i.e., $\gamma \!= \!0$, (\ref{eq:zeta}) reproduces the measurement-induced phase transition in Ref.~\onlinecite{jian2021syk}, where $\zeta \!>\! 0$ for $\tilde \mu \! < \! 1$ corresponds to a symmetry-breaking phase with volume-law entanglement and $\zeta \!=\! 0$ for $\tilde \mu \! \ge\! 1$ corresponds to a symmetric phase with area-law entanglement. 
The transition can be understood as symmetry restoration due to the strong measurement $ \mu \! > \! J$.
However, in the presence of the record loss error $\gamma \!> \!0$, the relative $C_4$ symmetry in (\ref{eq:action}) is explicitly broken. 
Moreover, one can see that $\zeta \! > \! \gamma $ for all values of $\tilde \mu$ as illustrated in Fig.~\ref{fig1}(a). 
Even at strong measurement $\tilde \mu \gg 1$, we have
$\zeta \!\approx\! \gamma [ 1 \!+\! \frac{(1-\gamma)(1 + \tilde U \gamma^2)}{\tilde \mu} ] \!>\! \gamma$. 
It indicates that the symmetry breaking transition is absent for nonzero errors.

We can proceed to evaluate the effective theory by treating $\gamma$ perturbatively. 
In terms of $\phi_1 \!=\! \delta G^{12} \!+\! \delta G^{34} $ and $ \phi_2 \!=\! \delta G^{14} \!+\! \delta G^{23} $~\cite{LR} that transforms as a vector $\vec \phi \!\equiv\! (\phi_1, \phi_2)$ under $C_4$ rotation, i.e., $\phi_1 \!\rightarrow\! -\phi_2$, $\phi_2 \!\rightarrow\! \phi_1$, the effective theory reads~\cite{suppl}
\bea \label{eq:Z4_model}
& \frac{I_{\text{eff}}}{N} = \frac12 \sum_{k} \int_{\Omega}  \left( \frac{\Omega^2}{\mu} + J(1-\cos k) \right) |\vec \phi_{k}(\Omega)|^2  
\nn \\
& +  \sum_x \int_{t} \left( \frac{\mu-J}2 \vec \phi_{x}^2 + \frac{\gamma\mu}2 \phi_1 + \frac{\mu}8 \vec \phi_{x}^4   - \frac{U}4 ( \phi_{1,x}^4 + \phi_{2,x}^4 ) \right).
\eea
where $\int_\Omega \!=\! \int \frac{d\Omega}{2\pi}$, $\int_t \!=\! \int dt$, and $\vec \phi_{k} \!=\! \frac1{\sqrt{L}}\sum_x \vec \phi_{x} e^{-i k x} $. It is apparent that $\gamma$ behaves like a magnetic field along $\phi_1$ direction, and breaks $C_4$ rotational symmetry explicitly.

\begin{figure}
    \centering
    \includegraphics[width=0.48 \textwidth]{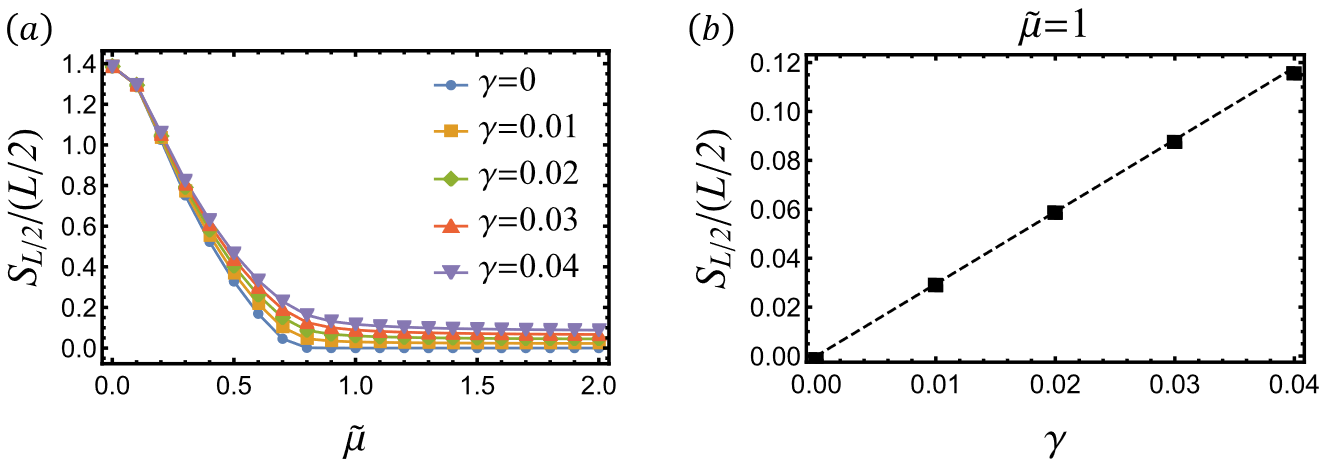}
    \caption{(a) The half-chain quasi entropy density of the steady state with and without record loss. (b) The entropy density at $\tilde \mu = 1$ is linear in $\gamma$. Here we take $\tilde{U}=0.4$, and to get the density we make a linear fit for sites $L=12,14,16,18$, and $TJ=L$.} \label{fig2}
\end{figure}

\B{\it Absence of measurement-induced phase transition.---} As shown in Ref.~\onlinecite{jian2021syk}, the swap operator in~(\ref{eq:entropy}) amounts to changing the boundary condition of the vector field $\vec \phi$ in the subsystem $A$, and the quasi entropy corresponds to the free energy difference between configurations with and without the change of the boundary condition.
To be concrete, we start from the thermofield double state (TFD) in the doubled Hilbert space~\cite{gu2017spread, penington2019replica, chen2020replica, jian2021syk} and divide the systems into two parts $A$ and $\bar A$. 
The quasi entropy at time $T/2$ is obtained by imposing twist operators at time $t\!=\!0$ and $t\!=\!T$ in the subsystem $A$, which require the boundary of $A$ has $(\phi_1\!=\!0,\phi_2\!>\!0)$ whereas that of $\bar A$ has $(\phi_1\!>\!0,\phi_2\!=\!0)$.
If the $C_4$ symmetry is broken by nonzero magnetic orders, the boundary condition will lead to different domains and will also induce the domain walls between them, as illustrated in Fig.~\ref{fig1}(b). 

For simplicity, we redefine the theory~(\ref{eq:Z4_model}) to be
\bea
    & \frac{I_\text{eff}}{N} \!=\! \int dt dx [ \frac12 (\partial \vec \phi)^2 + h \phi_1 + r \vec \phi^2 + \lambda \vec \phi^4 + \lambda' \left( \phi_1^4 + \phi_2^4 \right) ], \nn
\eea
with $ r \!=\! \frac{\mu (\mu - J)}2 $, $\lambda \!=\! \frac{\mu^{5/2}}{4\sqrt{2J} } $, $\lambda' \!=\! - \!2\tilde U \lambda $ and $ h \!=\! \gamma\frac{\mu^{7/4} J^{1/4}}{2^{5/4}} $. 
At the leading order, the free energy difference is given by the domain walls cost, $S_{A}^{(2)} = 2 N \sigma L_A$, where line tension can be obtained perturbatively for $r \le 0$,
\bea
\sigma & \approx  \frac{\pi \sqrt{-\lambda'}}8 ( \frac{-r}{\lambda + \lambda'} )^{3/2} + \frac{\pi -2}{(-2\lambda')^{1/2}} h.
\eea
The first term reproduces the line tension without errors in the measurement-induced phase transition~\cite{jian2021syk}, and the second term is independent of the tuning parameter $r$, which implies at $r\!=\!0$ the quasi entropy is still volume-law. On the other hand, for $r\! \gg \!0$, because the magnetic order pinned by the magnetic field points along $\phi_1$ direction in the bulk and pinned by the boundary condition points along $\phi_2$ direction at the boundary of subsystem $A$, they create a domain wall near the boundary with thickness given by the correlation length $\xi \!\propto\! r^{-1/2}$. 
Thus, the free energy cost is again linear in the subsystem length, i.e., $S_{A}^{(2)} \!\propto\! N h r^{-1/2} L_A$, and the quasi entropy is volume-law for $r\!>\!0$. 

In summary, the line tension is finite when $h\!>\!0$ and changes smoothly as a function of $r \!\propto\! \tilde \mu \!- \!1 $, so the measurement-induced phase transition is absent in the presence of quantum errors.
We confirm this conclusion with numerical results for the steady state quasi entropy density shown in Fig.~\ref{fig2}. 
Fig.~\ref{fig2}(a) shows that the transition disappears when $\gamma \!>\! 0$ because the entropy density is finite for all $\tilde \mu$. Fig.~\ref{fig2}(b) confirms the quasi entropy density at $\tilde \mu \!=\! 1$ is linear in $\gamma$.

\begin{figure}
    \centering
    \includegraphics[width=0.45 \textwidth]{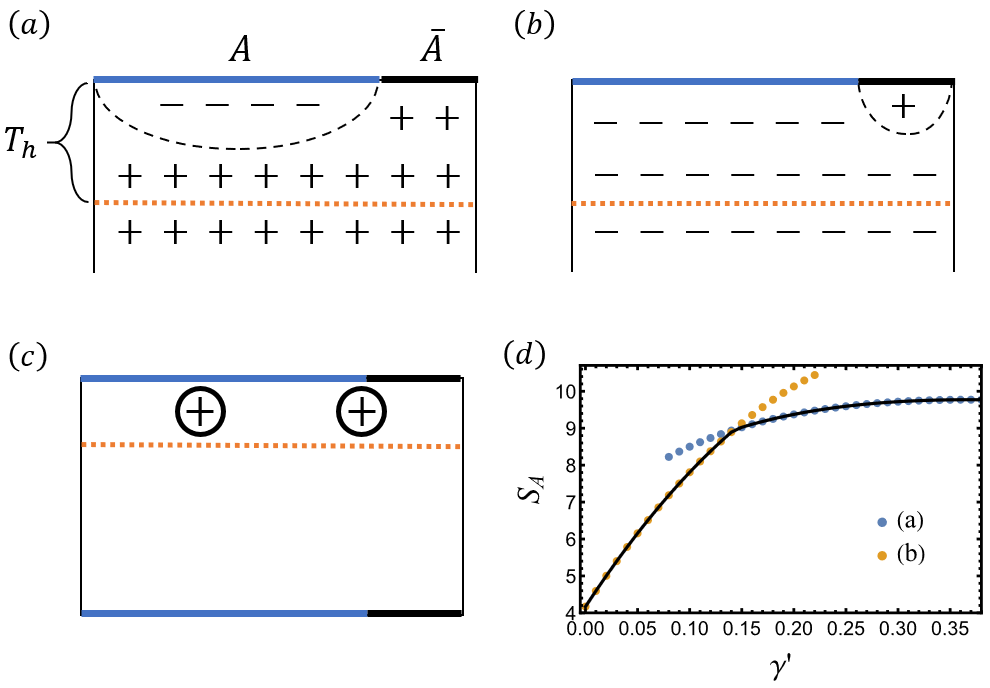}
    \caption{Two types of domain walls (a) and (b) in the presence of magnetic field near the boundary. 
    The magnetic field points toward $+$ direction. The orange dashed line is the boundary of the region with magnetic field, i.e., $T \!<\! t \!<\! T \! +\! T_h$. (c) shows the boundary condition and the magnetic field near the top boundary in the quasi entropy $S_A^{(2)}$ calculated in (d). The ``$+$" sign in the circle indicates the direction of the magnetic field. (d) The pinning transition of the subsystem quasi entropy of $L_A/L=2/3$ with $L=24$, $TJ=12$, $\tilde \mu = 0.6$, and $\tilde U = 0.4$. The entropy corresponding to type $(a)$ and $(b)$ are shown by the different colors. \label{fig3}}
    \label{fig3}
\end{figure}

\B{\it Pinning of domain walls.---} In an ordered magnet, the presence of an external magnetic field {\it near the boundary} can induce a first-order pinning transition of magnetic domains~\cite{pinning}. 
Consider the case where a field pointing toward the direction of $\phi_1$ (the “$+$" direction) exists in a region of width $T_h$ near the top boundary (i.e. the region $T\! <\! t \! <\! T \! + \!T_h \!$) with a “$+$" boundary condition $(\phi_1\!=\!0,\phi_2\!>\!0)$ for subsystem $\bar A$ and a “$-$" boundary condition $(\phi_1\!>\!0,\phi_2\!=\!0)$ for subsystem $A$. Note that $T_h$ is kept fixed when the thermodynamic limit is taken $ J T \sim L \rightarrow \infty$. 
Two types of domain walls are possible, one enclosing subsystem $A$ and the other $\bar A$ as shown in Fig.~\ref{fig3}(a) and (b), respectively. 
Since the magnetic field favors the “$+$" direction, it pushes the domain wall in Fig.~\ref{fig3}(a) away from the bulk but pulls the domain wall in Fig.~\ref{fig3}(b) into the bulk. 
To  leading order, the free energies of Fig.~\ref{fig3}(a) and Fig.~\ref{fig3}(b) are given respectively by~\cite{suppl}
\bea \label{eq:dowmain_wall}
    & F_a(x) =  N \sigma x, \quad F_b(x) = N \sigma x - N h T_h x .
\eea
where $x$ is the distance between the end points of domain walls at the boundary, and $\sigma$ ($h$) is the line tension (the magnetic field strength).

Below we assume $L_A \!>\! L_{\bar A}$ to observe a pinning transition.
At zero magnetic field, the configuration of Fig.~\ref{fig3}(b)  dominates over that of Fig.~\ref{fig3}(a) because $L_A \!>\! L_{\bar A}$. 
As the magnetic field increases, a pinning transition occurs when the two configurations have comparable free energies: $F_a(L_A) = F_b(L_{\bar A}) + N h T_h L  $, where in addition to the domain wall free energy in~(\ref{eq:dowmain_wall}) the energy cost of the disfavored domain in Fig.~\ref{fig3}(b) must be taken into account on the right-hand side.
This gives the transition point $h^\ast = \frac{2a-1}{a} \frac{\sigma}{T_h} $ ($a \!=\! \frac{L_A}{L}$), above which the dominant configuration switches to Fig.~\ref{fig3}(a).
Notice that in this calculation, we have implicitly assumed that the configurations in $0\!<\!t\!<\!T\!$ are the same. 

In the following, we also consider other domain walls near $t\!=\!0$ boundary. This only leads to a shift in the critical field for the pinning transition. 
The first-order pinning transition is manifest in the quasi entropy $S_A^{(2)}$ of the steady state (with initial state being TFD), where the corresponding boundary condition and emergent magnetic field are shown in Fig.~\ref{fig3}(c). 
Note that in this case the record loss occurs only near the boundary with a probability denoted by $\gamma'$ to distinguish from $\gamma$ which is used for the record loss throughout the circuit in the previous section. In the language of encoding introduced in next section, $\gamma$ ($\gamma'$) corresponds to record loss errors during (after) the encoding procedure.
A pinning transition between the two domain wall configurations occurs as one increases the probability $\gamma'$ as shown in Fig.~\ref{fig3}(d).

\begin{figure}
    \centering
    \includegraphics[width=0.48 \textwidth]{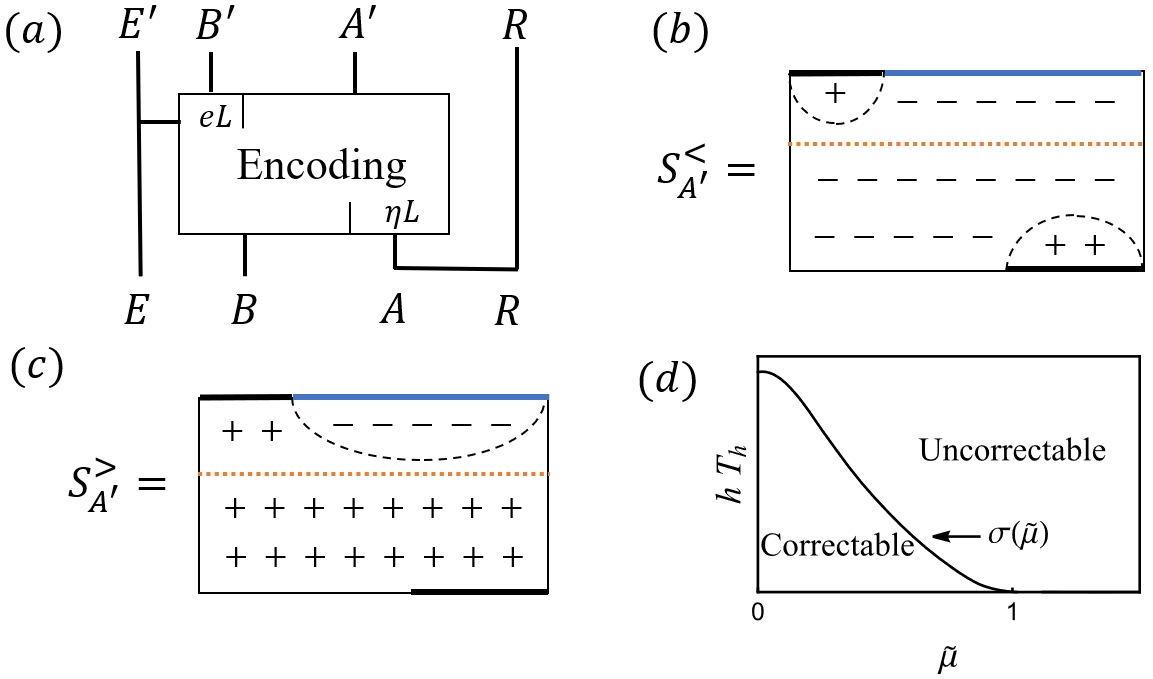}
    \caption{(a) Encoding process and possible quantum errors. See the main text for a detailed description. (b,c) Domain wall configurations corresponding to $S_{A'}^{(2)}$ before and after the pinning transition. The orange dashed line is the boundary of the region with magnetic field, i.e., $T \!<\! t \!<\! T \!+\! T_h$. (d) The error correction transition is given by the line tension without record loss. The correction to the line tension from the error is neglected.
     \label{fig4}}
\end{figure}

\B{\it Quantum error correction.---} The hybrid circuit can be understood as a quantum channel to generate an error correcting code for protecting quantum information~\cite{choi2020quantum,gullans2020dynamical,li2020statistical, li2021entanglement}.
A schematic plot of the encoding process and possible quantum errors are shown in Fig.~\ref{fig4}(a). 
To encode the information, one embeds it in region $A$ of the chain $|A| \!=\! \eta L$. By introducing a reference $R$, the initial state is given by a pure state in $AR$ tensor product with $B$, i.e., $ \rho = \rho_B \otimes | \psi_{AR} \rangle \langle \psi_{AR} |$. 
Applying a circuit of depth $T$ generates a code, $\mathcal E(\rho) \!=\! \frac{C_L \rho C_L^\dag}{\sqrt{\Tr(C_L^\dag C_L \rho)}}$, where $C_L$ denotes the circuit that acts on the system $AB$ but not on the reference $R$. 
Here, besides record loss, we also consider erasure errors that are commonly studied in the literature~\cite{gullans2020quantum}: the erasure error occurs in the region $B'$ of the chain $|B'| \!=\! e L$ at the boundary $t\!=\!T \!+ \! T_h$. This is to be compared with the record loss, which occurs with a probability $\gamma'$ near the boundary in the strip $[T\! , T \! + \!T_h]$. 
Notice that both errors happen at $t>T$ and we assume the encoding procedure is perfect in $0\!<\!t \!<\!T $.
It is useful to model the record loss as a coupling to an environment $E$.
Because we cannot access the information in region $E'B'$, the mutual information $I(R; E'B') \!=\! S_{R} \!+ \!S_{E'B'} \!- \!S_{E'B'R}$ measures the information diminished by errors~\cite{schumacher1996quantum}.
If $I(R;E'B') \!=\! 0$, it implies that the information is not lost and can be perfectly decoded.

At zero measurement rate, because the Brownian unitary dynamics can reach 2-design~\cite{onorati2017mixing, brandao2019models, frame_potential} in polynomial time, the decoupling theorem~\cite{abeyesinghe2009mother} applies and predicts the error threshold to be $1 \!-\! 2e_c \!-\! \eta \!=\! 0$.
It turns out the domain wall picture can reproduce the same result at zero measurement rate.
For finite measurement rate, although the decoupling theorem is not applicable, the domain wall picture and the associated pinning transition persist and provide an estimate of the quantum error threshold via the quasi-2 entropy. We assume the domain wall picture still holds for the von Neumann entropy and leave its verification for future study.

By introducing the environment $E$, the total pure wave function at time $T$ consists of four parts, $E',B',A'$ and $R$, so $S_{E'B'R}^{(2)} = S_{A'}^{(2)}$. 
Fig.~\ref{fig4}(b) and (c) show the configurations before and after the pinning transition in $S^{(2)}_{A'}$ respectively.  
Considering these two configurations, the entropy is
\bea
&	S^{(2)}_{A'} = - \log ( e^{-F_a((1-e)L)} + e^{- [F_b(eL) +  N h T_h L]} e^{- N \sigma \eta L } ). \nn 
\eea
In the second term inside the logarithm, the domain wall for the reference $R$ 
is included. [See Fig.~\ref{fig4}(b), where we have a domain wall ending at $t\!=\!0$ boundary.] 
The above equation determines the erasure error threshold $e_c$ in the presence of the record loss,
\bea \label{eq:threshold}
&	h T_h (1-e_c) = \sigma(1-2e_c-\eta).
\eea
Equivalently, if the erased portion is fixed, the transition is precisely the pinning transition at $h^\ast = \frac{1-2e-\eta}{1-e}\frac{\sigma}{T_h}$.

It is worth noting that there is also a similar pinning transition in $S_{E'B'}^{(2)}$. Including pinning transitions in both $S_{A'}^{(2)}$ and $S_{E'B'}^{(2)}$, the mutual information reads
\bea
&&	I^{(2)}(R;E'B') = \nn \\
&& \qquad \qquad \begin{cases}
		0                                       & e \!<\! e_c \\
		N[\sigma(\eta + 2e -1) + h T_h (1-e)]L & e_c \!<\! e \! < e_* \\
		2 N \sigma \eta L                       & e_* \!<\! e
	\end{cases}, \nn
\eea
where $e_\ast$ given by $h T_h (1-e_\ast) \!=\! \sigma(1 \!-\! 2e_\ast \!+\! \eta )$ is from the pinning transition in $S_{E'R'}^{(2)}$.
If no erasure error occurs, $\sigma \!=\! h T_h$ according to~(\ref{eq:threshold}), which indicates an error correction transition due to the record loss.
Neglecting the correction to the line tension from the magnetic field when the record loss error is small, the threshold is then determined by the line tension in zero magnetic field as shown in Fig.~\ref{fig4}(d).

\B{\it Concluding remarks.---} To conclude, with a concrete solvable model we show that the record loss error effectively generates a coupling between two Keldysh contours within each replica. 
Such a coupling explicitly breaks the permutation symmetry among the forward (backward) contours in different replica.
We expect this result to be valid in more general errors because tracing out an environment that is coupled to the system will generate the inter-Keldysh coupling within each replica. 

Remarkably, the domain wall picture of quantum error correction can also be generalized to the Hayden-Preskill protocol~\cite{hayden2007black}, where the information of part $B$ is collected before the encoding process. This can be interpreted as the collected Hawking radiation from the early black hole. 
A direct application of the domain wall picture leads to the error threshold $h T_h (1-e) \!=\! 2\sigma(1-e-\eta)$~\cite{suppl}.
It would be interesting to explore the effect of the quantum error on the decoding process~\cite{yoshida2017efficient}.

\B{\it Acknowledgement.---} SKJ and BGS are supported by the Simons Foundation via the It From Qubit Collaboration. The work of BGS is also supported in part by the AFOSR under grant number FA9550-19-1-0360. PZ acknowledges support from the Walter Burke Institute for Theoretical Physics at Caltech. CL is supported by the NSF CMMT program under Grants No. DMR-1818533. We acknowledge the University of Maryland High Performance Computing Cluster (HPCC).

\bibliography{reference.bib}

\begin{thebibliography}{58}
\bibitem [{EE()}]{EE}%
  \BibitemOpen
  \href@noop {} {}\bibinfo {note} {Technically, the calculable quantity is not
  entanglement entropy but the quasi entropy defined in~(\ref{eq:entropy}). We
  expect that the entanglement entropy also has a transition since the quasi
  entropy is believed to be a useful proxy of entanglement
  entropy.}\BibitemShut {Stop}%
\bibitem [{LR()}]{LR}%
  \BibitemOpen
  \href@noop {} {}\bibinfo {note} {In this case we have $\delta G^{12}_{RR} =
  \delta G^{12}_{LL}$, $\delta G^{34}_{RR} = \delta G^{34}_{LL}$, $\delta
  G^{14}_{RR} = \delta G^{14}_{LL}$, $\delta G^{23}_{RR} = \delta
  G^{23}_{LL}$~\cite{suppl}, so we omit the subscript of the left and right
  chains.}\BibitemShut {Stop}%
\bibitem [{fra()}]{frame_potential}%
  \BibitemOpen
  \href@noop {} {}\bibinfo {note} {We also calculate the frame potential of the
  Brownian SYK chain~(\ref{eq:LRhamiltonian}) without monitoring~\cite{suppl},
  which shows that the Brownian SYK chain can achieve the 2
  design.}\BibitemShut {Stop}%
\end{thebibliography}

\newpage
\onecolumngrid

\section{Supplemental Material for ``Quantum error as an emergent magnetic field"}

\setcounter{secnumdepth}{3}
\setcounter{equation}{0}
\setcounter{figure}{0}
\renewcommand{\theequation}{S\arabic{equation}}
\renewcommand{\thefigure}{S\arabic{figure}}
\newcommand\Scite[1]{[S\citealp{#1}]}
\makeatletter \renewcommand\@biblabel[1]{[S#1]} \makeatother

\section{Derivation of the large-$N$ action and the saddle-point equation of the monitored system}

The derivation of the large-$N$ action of the Brownian SYK model is a standard one~\cite{saad2018semiclassical}, so we focus on the effect of the measurement. 
As mentioned in the main text, the system is under continuously monitoring given by the measurement operators for the $i$-th Majorana fermion of left and right chains at site $x$,
\bea
\{ M_1, M_2\} = \left\{ \Big( \frac{1+\sqrt{1-4s^2}}{2} \Big)^{1/2} - \Big( \frac{1 - \sqrt{1-4s^2}}{2} \Big)^{1/2} \left(\frac12 + i \psi_{x,L,i}\psi_{x,R,i} \right)    ,s \left(\psi_{x,L,i} + i \psi_{x,R,i} \right) \right\},
\eea
where $0< s < \frac12$ is the strength of measurement. 
During a time step, the measurement with a probability $p$ leads to the following operator for the two replicas,
\bea \label{eq:evolution}
	&& (1-p) \mathbb I + p (1 - \gamma) \sum_i M_i \otimes M_i^\dag \otimes  M_i \otimes M_i^\dag + p \gamma \sum_i  M_i \otimes M_i^\dag \otimes \sum_j M_j \otimes M_j^\dag \\
	&& = 1 - p + p (1 - \gamma) \Big(1 - s^2  \sum_{\alpha=1}^4 (\frac12 + i \psi_{x,L,i}^\alpha \psi_{x,R,i}^\alpha ) \Big) + p \gamma K_{43} K_{21},
\eea
where $K_{kl}$ in the second line is defined as
\bea
	K_{kl} = 1- s^2 \sum_{\alpha=k}^l(\frac12 + i \psi_{x,L,i}^\alpha \psi_{x,R,i}^\alpha) +s^2 (\psi_{x,L,i}^k \psi_{x,L,i}^l + \psi_{x,R,i}^k \psi_{x,R,i}^l -i \psi_{x,L,i}^k \psi_{x,R,i}^l + i \psi_{x,R,i}^k \psi_{x,L,i}^l).
\eea
and the derivation is kept up to the $O(s^2)$ order. $\gamma$ is the probability of losing the measurement outcome when a measurement is implemented.
We perform the measurement for every Majorana species $i$ at every site $x$, and the result is the following effective action
\bea \label{eq:measurement}
	\exp \delta t \left( - \frac{\mu}2 \sum_\alpha i G_{x, LR}^{\alpha\alpha} - \frac{\gamma \mu}2 \big[ M^{12} + M^{34} \big]  \right).
\eea
where $M^{\alpha\beta} =  G_{LL}^{\alpha\beta} + G_{RR}^{\alpha\beta} - i  G_{LR}^{\alpha\beta} + i G_{RL}^{\alpha\beta}$, and $\mu = p s^2 /\delta t$ is the measurement rate. This is (6) in the main text.
Integrating out the Gaussian variables, the large-$N$ action is
\bea 
- \frac{I}N &=& \frac12 \Tr \log \left( (-1)^{\alpha+1} \delta^{\alpha\beta} \delta_{ab} \partial_t - \Sigma_{ab,x}^{\alpha\beta} \right) + \iint - \frac12 \Sigma_{ab,x}^{\alpha\beta}(t,t') G_{ab,x}^{\alpha\beta}(t,t') 	\nn \\
&& + \iint \delta(t-t') \Big[ - \frac{(-1)^{\alpha + \beta}}{4} \delta_{ab} \left( J G_{ab,x}^{\alpha\beta} G_{ab,x+1}^{\alpha\beta} + \frac{U}{2q} (2G_{ab,x}^{\alpha\beta})^{q} \right) \nn \\
 && - \frac{\mu}2 \sum_\alpha i G_{x, LR}^{\alpha\alpha} - \frac{\gamma \mu}2 \big[ M^{12} + M^{34} \big] \Big], 
\eea
where $\alpha, \beta = 1,...,4$ denote the four contours.
The summations over $x$, $a,b$ and $\alpha, \beta$ are implicit.
$\Sigma_{ab,x}^{\alpha\beta}(t,t')$ is the self-energy field introduced to enforce $ G_{ab,x}^{\alpha\beta}(t,t') = \frac1N \sum_j \psi_{x,a,j}^\alpha(t) \psi_{x,b,j}^\beta(t')$. The saddle point equation reads
\bea
 [G_x^{-1}]^{\alpha\beta}_{ab} &=& (-1)^{\alpha+1} \delta^{\alpha\beta} \delta_{ab}\partial_t - \Sigma_{ab,x}^{\alpha\beta}, \\
\Sigma_{ab,x}^{\alpha\beta}(t,t') &=& \delta(t-t') \Big[ \frac{- (-1)^{\alpha+\beta} \delta_{ab}}{2} \left( J (G_{ab,x-1}^{\alpha\beta} + G_{ab,x+1}^{\alpha\beta})  + U(2 G_{ab,x}^{\alpha\beta})^{q-1} \right) \nn \\
&&   - i \frac{\mu}2 \bar\delta_{LR}^{\alpha\alpha} - \big(\frac{\gamma \mu}2 \bar\delta_{aa}^{12}- i \frac{\gamma \mu}2 (\bar\delta_{LR}^{12} - \bar\delta_{RL}^{12})+\frac{\gamma \mu}2 \bar\delta_{aa}^{34}- i \frac{\gamma \mu}2 (\bar\delta_{LR}^{34} - \bar\delta_{RL}^{34}) \big) \Big].
\eea
where we define a short-hand notation $ \bar \delta_{cd}^{\gamma\delta} \equiv \delta^{ac} \delta^{bd} \delta_{\alpha\gamma} \delta_{\beta\delta} - \delta^{bc} \delta^{ad} \delta_{\beta\gamma} \delta_{\alpha\delta}$.

\section{Derivation of the effective action~(10)} 

We proceed to look at the effective theory by treating $\gamma$ perturbatively. 
Without the magnetic field, the symmetric solution reads
\bea \label{eq:SYKq_saddle2}
    \bar G(t_1,t_2) &=& \frac{e^{- \frac{\mu}2 |t_{12}|}}{2} 
    \left( \ba{cccc} \sgn(t_{12}) & 0 & i & 0 \\ 
    0 & - \sgn(t_{12}) & 0 &   i \\ 
    -i & 0 & \sgn(t_{12}) & 0 \\
    0 & -i & 0 & - \sgn(t_{12}) 
    \ea \right).
\eea
where the basis of the matrix is $(L1,L2,R1,R2)$. And the solution is the same for contour $3$ and $4$. 
We consider the fluctuation away from the symmetric saddle-point solution~(\ref{eq:SYKq_saddle2}), so that the effect of the record loss is reflected in the effective action. 
The kernel of $\delta \Sigma$ from expanding the trace log term can again be brought into decoupled sectors, and because they serve as an order parameter we focus on the components $(\delta \Sigma^{12}_{aa},\delta \Sigma^{34}_{aa},\delta \Sigma^{14}_{aa},\delta \Sigma^{23}_{aa})$ whose kernel is given by
\bea \label{eq:kernel}
	K = \frac{\mu}{4(\mu^2 + \Omega^2)} 1_{4\times 4} \otimes \left( \ba{cccc} 1 & 1 \\ 1 & 1 \ea \right),
\eea
where the first matrix is in the basis of these four components and the second is in the basis of $L$ and $R$ chains.
It is apparent that there are four zero modes and integrating them out will lead to the following constraints,
\bea
	\delta G^{12}_{RR} = \delta G^{12}_{LL},\quad \delta G^{34}_{RR} = \delta G^{34}_{LL}, \quad \delta G^{14}_{RR} = \delta G^{14}_{LL}, \quad \delta G^{23}_{RR} = \delta G^{23}_{LL}.
\eea
Thus, there are four independent fields left $(\delta G^{12},\delta G^{34},\delta G^{14},\delta G^{23})$ where we suppress the subscript.

Now it is a straightforward task to integrate out the rest fluctuations with nonzero kernel in~(\ref{eq:kernel}). 
After identifying $ \phi_1 = \delta G^{12} + \delta G^{34} $ and $ \phi_2 = \delta G^{14} + \delta G^{23} $ as the order parameter which under the $C_4$ rotation transforms like a vector, i.e., $(\phi_1, \phi_2) \rightarrow (\phi_2, - \phi_1)$, the effective theory reads
\bea \label{eq:Z4_model}
\frac{I_{\text{eff}}}{N} &=& \frac12 \sum_{i=1,2; k} \int_{\Omega}  \left( \frac{\Omega^2}{\mu} + J(1-\cos k) \right) |\phi_{i,k}(\Omega)|^2  
 +  \sum_x \int_{t} \left( \frac{\mu-J}2 \vec \phi_{x}^2 + \frac{\gamma\mu}2 \phi_1 + \frac{\mu}8 \vec \phi_{x}^4   - \frac{U}4 ( \phi_{1,x}^4 + \phi_{2,x}^4 ) \right).
\eea
which is (10) in the main text.

\section{Derivation of the frame potential in the Brownian SYK circuit}

For the Brownian SYK model defined in~(1), $\Tr[U V^\dag]$ can be brought to $\Tr[U]$ by just redefining the Hamiltonian because there is no memory and the mean of the coupling~(2) is zero. 
After this trick, the frame potential becomes 
\bea
F_{bSYK}^{(m)}(t)=\overline{ |\Tr U(t)|^{2m} }=\overline{ [\Tr U(t)]^m [\Tr U(t)^\ast]^{m} }.
\eea
where the overline indicates the average over the Brownian variable. Without the monitoring, because there is no coupling between the left and right chains, we can consider only one of the chains. 
After integrating over the random variable, the frame potential reads
\bea
&& \overline{ [\Tr U(t)]^m [\Tr U(t)^\ast]^m} \\
&=& \int DG D\Sigma \exp N  \Big[ \frac12 \log \det [\partial_t + \Sigma_x] + \int \Sigma_{LR,x}^{\alpha\beta} G_{LR,x}^{\alpha\beta} + \frac12  \int dt \Big[ -J\int (G_{LR,x}^{\alpha\beta})^2 + i^q \frac{U}{2q} (2 G_{LR,x}^{\alpha\beta})^q - mL\Big(\frac{J}4 + \frac{U}{2q} \Big) \Big]. \nn\\
\eea
where $G_{ab,x}^{\alpha\beta}(t) = \frac1N \sum_i \psi_{x,a,i}^\alpha(t) \psi_{x,b,i}^\beta(t)$, $a = L, R$ and $\alpha=1,..,m$.
Note that in this section $L$ and $R$ does not denote two chains in the main text, but the two contours, one forward and one backward, in one of the $m$ replicas, and $\alpha$ is the replica index.
For our purpose, it is enough to assume $G, \Sigma$ fields are constants independent of the site. 
In this case, the $\log \det$ part can be mapped to a partition function of Majorana fermions with Hamiltonian~\cite{saad2018semiclassical} $ H =\frac12 \left( \ba{cccc} 0 &  \Sigma_{LR} \\ - \Sigma_{LR}^T & 0 \ea \right)$. We assume that when the replica $\alpha$ pairs with the replica $\beta$, i.e., $G^{\alpha\beta}_{LR} = \pm \frac{i}2$, all other pairing with replica $\alpha$ is zero $G^{\alpha\alpha'}_{LR}=0$, $\alpha' \ne \beta$. 
Thus, we can get the $\log \det$,
\bea
\frac12 \log \det [\partial_t + \Sigma] = \log \int D\chi \exp - \int dt \chi^\alpha \Big( \frac12 \partial_t \delta^{\alpha\beta} + H \Big) \chi^\beta = \log 2\cos\frac{\Sigma_{LR}^{\alpha\beta} t}2.
\eea
which results in the effective action,
\bea\label{Action1}
\overline{ [\Tr U(t)]^m [\Tr U(t)^\ast]^m} = e^{ N L \sum_{\alpha\beta} \Big[ \log 2\cos\frac{\Sigma_{LR}^{\alpha\beta} t}2+ \Sigma_{LR}^{\alpha\beta} G_{LR}^{\alpha\beta} t + \frac{Jt}8 (2i G_{LR}^{\alpha\beta})^2 + \frac{U t}{4q} (2i G_{LR}^{\alpha\beta})^q   \Big]} e^{ - \frac{m N L}2\Big(\frac{J}4 + \frac{U}{2q}\Big) t}.
\eea
For a fixed $\alpha\beta$ the equation of motion is
\bea
G_{LR} &=& \frac12 \tan \frac{\Sigma_{LR} t}2,  \\
\Sigma_{LR} &=& J G_{LR}^{\alpha\beta} - \frac{i U}{2} (2 i G_{LR})^{q-1}.
\eea
The trivial solution is $G_{LR} = \Sigma_{LR} = 0$ and the wormhole solution at $t \rightarrow \infty$ is $G_{LR} = \pm \frac{i}2$, $\Sigma_{LR} = \pm i \frac{J+U}2$. 
Note that the plus and minus choice is a gauge symmetry, because one can refine $\psi_{x,a,i}^\alpha \rightarrow -\psi_{x,a,i}^\alpha $ without changing anything. For the trivial solution, the first term in~(\ref{Action1}) leads to a constant $2^{NL}$. While for the wormhole solution, the factor term in~(\ref{Action1}) is
\bea
e^{N L (\log 2 \cosh \frac{(J+U) t}{4} - \frac{(J+U) t}{4} + \frac{Jt}{8}+\frac{U t}{4q} )}\rightarrow e^{N Lt(\frac{J}{8}+\frac{U}{4q} )},
\eea
leading to a nontrivial contribution. 
The remaining step is to count how many different wormhole solutions can exist, so we get
\bea
F_{bSYK}^{(m)}(t) &=& e^{- m N Lt(\frac{J}{8}+\frac{U}{4q} )} \sum_{k=0}^m \frac{m!^2}{k! (m-k)!^2} e^{(m-k)NL \log 2} e^{k N Lt(\frac{J}{8}+\frac{U}{4q})} \\
&=& (-1)^m U\left( -m,1, - e^{NL(\log2 - (\frac{J}{8}+ \frac{U}{4q})t)}\right)  \\
&\rightarrow& \begin{cases}
	2^{mNL }(1+ \mathcal{O}(2^{-NL}))\quad & t \rightarrow 0 \\
	m! \quad & t \rightarrow \infty
\end{cases}.
\eea
where $U(a,b,z)$ is a confluent hypergeometric function.
At time zero it is the dimension of Hilbert space of $2m$ Brownian SYK chains (each with $NL$ Majorana), while at time $t \rightarrow \infty$, it counts the number of $m$-connected wormholes.
The frame potential is bounded from below by $F^{(m)} \ge F^{(m)}_\text{Haar} = m!$.
The equality holds if and only if the ensemble achieves $m$-design~\cite{roberts2017chaos}.
Then our results show that the Brownian SYK chain can achieve $m$-design for long enough time, and this is consistent with the conclusion that a Brownian Hamiltonian can achieve $m$-design for polynomial time~\cite{brandao2019models}.

\section{Derivation of the free energy of domain walls in~(12)}

\begin{figure}
    \centering
    \includegraphics[width=0.3 \textwidth]{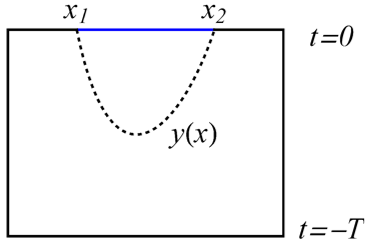}
    \caption{A schematic plot of the domain wall ending at $[x_1, x_2]$ on the $t=0$ boundary. The dashed curve indicates the domain wall. The blue line indicates the boundary with twisted boundary conditions.}
    \label{fig_s1}
\end{figure}

As we discussed in the main text, the entanglement entropy can be mapped to the free energy cost of domain walls induced by the twisted boundary condition. 
A schematic plot of the domain wall induced by the twisted boundary condition within $[x_1,x_2]$ at $t=0$ is shown in Fig.~\ref{fig_s1}.
The domain wall is given by the function $y(x) \in [-T, 0]$, and its free energy reads
\bea
    S_A = - \log \int_{y(x_1) = y(x_2) = 0} D y e^{- \int dx \left( \sigma \sqrt{1 + (\partial_x y)^2} + V(y) \right)}, 
\eea
where $\sigma$ is the surface tension and $V(y)$ will be specified below. 
Let us assume the fluctuation is small so that we can approximate $\sqrt{1 + (\partial_x y)^2} \approx 1 + \frac12  (\partial_x y)^2 $.
Then we can calculate the entanglement entropy by mapping it to a transition amplitude of a non-relativistic particle with mass $\frac1{2\sigma}$ in a potential
\bea
\int_{y(x_1)}^{y(x_2)} D y e^{- \int dx \left(  \frac{\sigma}2(\partial_x y)^2 +V(y) \right)} = \langle y(x_2) | e^{- H (x_2 - x_1)} | y(x_1)\rangle,
\eea
with $H = \frac{1}{2\sigma} p_y^2 +V(y)$, where $y $ and $p_y$ are conjugate variables, $[y, p_y] = i$.

In the quantum mechanical picture, we use $y$ to denote the depth of the circuit (or the evolution time of the circuit). 
We will be interested in two cases, $V_+(y)$ and $V_-(y)$, where the magnetic field exists near the boundary and the potential favors and disfavors the domain wall,
\bea
V_+(y) = \begin{cases} 
	\infty & 0<y \\
	- h y &-T_h<y<0 \\
	h T_h & -T_h - T < y<-T_h  \\
	\infty & y<-T_h - T
\end{cases}, \quad 
V_-(y) = \begin{cases} 
	\infty & 0<y \\
	h y &-T_h<y<0 \\
	- h T_h & -T_h - T < y<-T_h  \\
	\infty & y<-T_h - T
\end{cases} .
\eea

\begin{figure}
	\centering
	\includegraphics[width=0.7\textwidth]{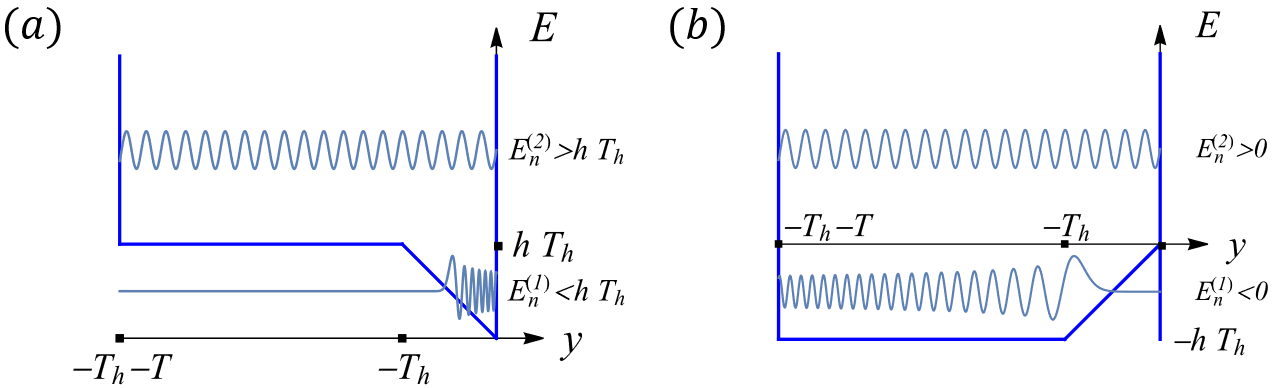}
	\caption{Illustration of two kinds of the eigen-energy in the WKB method in the two potentials $V_+(y)$ (a) and $V_-(y)$ (b).\label{fig:WKB}}
\end{figure}

The potential $V_+(y)$ corresponds to the potential felt by the domain wall shown in Fig.~\ref{fig:WKB}(a), while the second potential $V_-(y)$ corresponds to the potential felt by the domain wall shown in Fig.~\ref{fig:WKB}(b).
We use the WKB method to calculate the transition amplitude. 
The approximate energy is given by the formula 
\bea
\int_{y_1}^{y_2} dy \sqrt{2\sigma (E_n-V(y))} = \left(n - \frac12 \right) \pi ,
\eea
where $y_1<y_2$ are two turning points.
There are two types of eigenstates, with energy greater or less than the linear potential, as shown in Fig.~\ref{fig:WKB}.
For the potential $V_+(y)$ the first type of eigenstate is trapped in the linear potential and presents the domain wall located in the region with magnetic field.
It has turning point $ [- E^{(1)}_n/h, 0]$, and its eigen-energy is given by
\bea
&& \int_{- E^{(1)}_n/h}^{0} dy \sqrt{2\sigma(E^{(1)}_n + hy)} =  \left(n - \frac12 \right) \pi, \\
&& E^{(1)}_n = \left( \frac{h^2}{2\sigma} \right)^{1/3} \left( \frac{3\pi}{4}(2n-1) \right)^{2/3}, \quad 1 \le n \le n_* ,\quad E_{n_*}^{(1)} = h T_h.
\eea
It is apparent that $n_*$ is independent of $L$.
The second type of eigenstate extends to the flat potential and presents the domain wall that form in the region without magnetic field.
It has turning point $[-T-T_h, 0 ]$, and its eigenenergy is determined by
\bea
&& \int_{-T_h}^{0} dy \sqrt{2\sigma(E^{(2)}_n + h y)} + \int_{-T_h}^{-T_h-T} dy \sqrt{2\sigma(E^{(2)}_n - h T_h)} \\
&& = \frac23 \left(\frac{2\sigma}{h^2} \right)^{1/2} \left((E^{(2)}_n)^{3/2} - (E^{(2)}_n - h T_h)^{3/2}  \right) + \sqrt{2\sigma(E^{(2)}_n - h T_h)} T = n  \pi, \\
&& E^{(2)}_n = h T_h + \frac1{2\sigma} \left( \frac{n\pi}T \right)^{2}, \quad E^{(2)}_n \gg hT_h,
\eea
where in the second line we have replaced $n + \frac12 $ by $n$ because we know that the turning point is actually a node, and in the last line we make the approximation $E^{(2)}_n \gg hT_h$.
The transition amplitude reads
\bea
&& \langle y(x_2) | e^{- H_+ (x_2 - x_1)} | y(x_1)\rangle = \sum_{n=1}^\infty  e^{- E_n (x_2 - x_1)} \psi^\ast_n(-\epsilon) \psi_n(-\epsilon) \nn \\
&=& \sum_{n=1}^{n_\ast}  e^{- E^{(1)}_n (x_2 - x_1)} \psi^{(1)\ast}_n(-\epsilon) \psi^{(1)}_n(-\epsilon) + \sum_{n=1}^{\infty}  e^{- E^{(2)}_n (x_2 - x_1)} \psi^{(2)\ast}_n(-\epsilon) \psi^{(2)}_n(-\epsilon),\\
&\approx& e^{- \left( \frac{3\pi}{4} \right)^{2/3} \left( \frac{h^2}{2\sigma} \right)^{1/3}  x_{12}} \sum_{n=1}^{n_\ast} \psi^{(1)\ast}_n(-\epsilon) \psi^{(1)}_n(-\epsilon) + e^{-hT_h x_{12}} \sqrt{\frac{2}{\pi}} \epsilon^2 (2\sigma)^{3/2} x_{12}^{-3/2}, \label{eq:fisrt_wall}
\eea
where in the last step, we approximate the wave function $\psi^{(2)}_n(y)$ by the eigenstate of a infinite potential box $\psi^{(2)}_n(y) = \sqrt{\frac{2}{T}} \sin \frac{n\pi y}T$, and extends the summation by an integral.
Because we use the large-$N$ approximation to get the domain wall free energy, the $N$ should be restored by taking $\sigma \rightarrow N \sigma$ and $h \rightarrow N h$. 
Thus the first term in~(\ref{eq:fisrt_wall}) dominates in the large-$N$ limit, and the transition amplitude reads,
\bea
\langle y(x_2) | e^{- H_+ (x_2 - x_1)} | y(x_1)\rangle &\approx& e^{- N^{1/3} \left( \frac{3\pi}{4} \right)^{2/3} \left( \frac{h^2}{2\sigma} \right)^{1/3}  x_{12}} \sum_{n=1}^{n_\ast} \psi^{(1)\ast}_n(-\epsilon) \psi^{(1)}_n(-\epsilon) \sim e^{- N^{1/3} \left( \frac{3\pi}{4} \right)^{2/3} \left( \frac{h^2}{2\sigma} \right)^{1/3}  x_{12}}.
\eea
The domain wall of the first type has the free energy
\bea \label{eq:F_+}
	F_a(x_{12}) = N \sigma x_{12} + N^{1/3} \left( \frac{3\pi}{4} \right)^{2/3} \left( \frac{h^2}{2\sigma} \right)^{1/3}  x_{12} \approx N \sigma x_{12} .
\eea

Now a similar analysis for the potential $V_-(y)$ can be lay out.
For the potential $V_-(y)$ the first type of eigenstate is trapped in the linear potential and presents the domain wall located in the region with magnetic field.
It has turning point $ [-T-T_h, E^{(1)}/h]$, and its eigen-energy is given by
\bea
&& \int_{-T-T_h}^{-T_h} dy \sqrt{2\sigma(E^{(1)}_n + h T_h)} +  \int_{-T_h}^{ E^{(1)}_n/h} dy \sqrt{2\sigma(E^{(1)}_n - hy)} \nn \\
&=&  T \sqrt{2\sigma(E^{(1)}_n + h T_h)} + \frac23 \left(\frac{2\sigma}{h^2} \right)^{1/2} \left(h T_h + E_n^{(1)} \right)^{3/2}  = \left(n - \frac12 \right) \pi , \\
&& E^{(1)}_n = - h T_h + f \left( (n- \frac12)\pi \right), \quad 1 \le n \le n_* ,\quad E_{n_*}^{(1)} = 0.
\eea
where $f$ is the solution to $T \sqrt{2\sigma f} + \frac23 \left(\frac{2\sigma}{h^2} \right)^{1/2} f^{3/2}  = \left(n - \frac12 \right) \pi$. 
The second type of eigenstate extends to the flat potential and presents the domain wall that form in the region {\it without} magnetic field.
It has turning point $[-T-T_h, 0 ]$, and its eigenenergy is determined by
\bea
&& \int_{-T_h}^{0} dy \sqrt{2\sigma(E^{(2)}_n - hy)} + \int_{-T_h}^{-T_h-T} dy \sqrt{2\sigma(E^{(2)}_n + h T_h)} \nn \\
&& = \frac23 \left(\frac{2\sigma}{h^2} \right)^{1/2} \left((E^{(2)}_n + h T_h)^{3/2} - (E^{(2)}_n)^{3/2}\right) + \sqrt{2\sigma(E^{(2)}_n + h T_h)} T = n  \pi, \\
&& E^{(2)}_n = -h T_h + \frac1{2\sigma} \left( \frac{n\pi}T \right)^{2}, \quad E^{(2)}_n \gg hT_h,
\eea
where in the second line we have replaced $n - \frac12 $ by $n$ for the same reason, and in the last line we make the approximation $E^{(2)}_n \gg hT_h$.
The transition amplitude reads
\bea
&& \langle y(x_2) | e^{- H_- (x_2 - x_1)} | y(x_1)\rangle = \sum_{n=1}^\infty  e^{- E_n (x_2 - x_1)} \psi^\ast_n(-\epsilon) \psi_n(-\epsilon) \\
&=& \sum_{n=1}^{n_\ast}  e^{- E^{(1)}_n (x_2 - x_1)} \psi^{(1)\ast}_n(-\epsilon) \psi^{(1)}_n(-\epsilon) + \sum_{n=1}^{\infty}  e^{- E^{(2)}_n (x_2 - x_1)} \psi^{(2)\ast}_n(-\epsilon) \psi^{(2)}_n(-\epsilon),\\
&\approx&  e^{h T_h x_{12}} \left( e^{- f((n- \frac12 )\pi)  x_{12}} \sum_{n=1}^{n_\ast} \psi^{(1)\ast}_n(-\epsilon) \psi^{(1)}_n(-\epsilon) + \sqrt{\frac{2}{\pi}} \epsilon^2 (2\sigma)^{3/2} x_{12}^{-3/2} \right),
\eea
where in the last step, we approximate the wave function $\psi^{(2)}_n(y)$ by the eigenstate of a infinite potential box $\psi^{(2)}_n(y) = \sqrt{\frac{2}{T}} \sin \frac{n\pi y}T$, and extends the summation by an integral. 
Then after restoring the factor of $N$, we have the transition amplitude with potential $V_-(y)$,
\bea
\langle y(x_2) | e^{- H_- (x_2 - x_1)} | y(x_1)\rangle &\sim& e^{N h T_hx_{12}}.
\eea
The domain wall of the second type has the free energy
\bea \label{eq:F_-}
F_b(x_{12}) = N \sigma x_{12} - N h T_h x_{12} .
\eea
(\ref{eq:F_+}) and (\ref{eq:F_-}) give (12) in the main text.

\section{Domain wall picture in the Hayden-Preskill protocol} 

\begin{figure}
    \centering
    \includegraphics[width=0.3\textwidth]{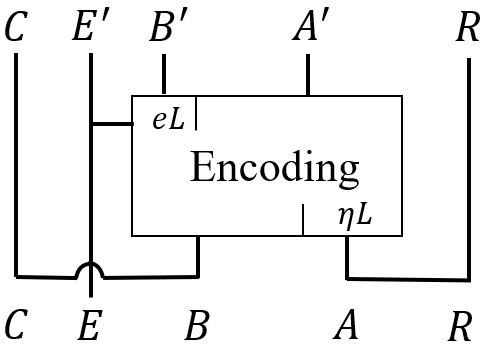}
    \caption{The schematic plot of the encoding process. The information is encoded in $A$. And we collect all the information in $B$. To model the encoded information, we introduce a reference $R$ such that $A$ and $R$ are maximally entangled. We also introduce part $C$ that is maximally entangled with $B$ to model the collection of information in $B$. $B'$ denotes the erasure error and $E$ is introduced to model the record loss error.}
    \label{fig:s3}
\end{figure}

In the main text we discuss the encoding process where no information about the $B$ part of the circuit is known apriori. 
Here we generalize the discussion to the case where we have the information about the state in $B$ before we encode the information.
In the Hayden-Preskill thought experiment, $B$ is the early radiation of the black hole collected by Bob~\cite{hayden2007black}.
The encoding process is schematically shown in Fig.~\ref{fig:s3}. 
Comparing to Fig.4(a) in the main text, the difference is we have collected the information of part $B$, which is modeled by a maximally entangled part $C$.
We are still interested in the mutual information between $E'B'$ and $R$, i.e.,
\bea
    I(E'B';R) = S_R + S_{E'B'} - S_{E'B'R}.
\eea
First consider the case with zero measurement rate (thus no record loss error). In this case the decoupling theorem states that the deviation between the density matrix $\rho_{B'R}$ and the tensor product density matrix $\rho_{B'} \otimes \rho_R$ is bounded by $2^{-N (1-e-\eta) L}$~\cite{abeyesinghe2009mother}. 
Thus the error threshold is $1-e_c-\eta=0$.

\begin{figure}
    \centering
    \includegraphics[width=0.7\textwidth]{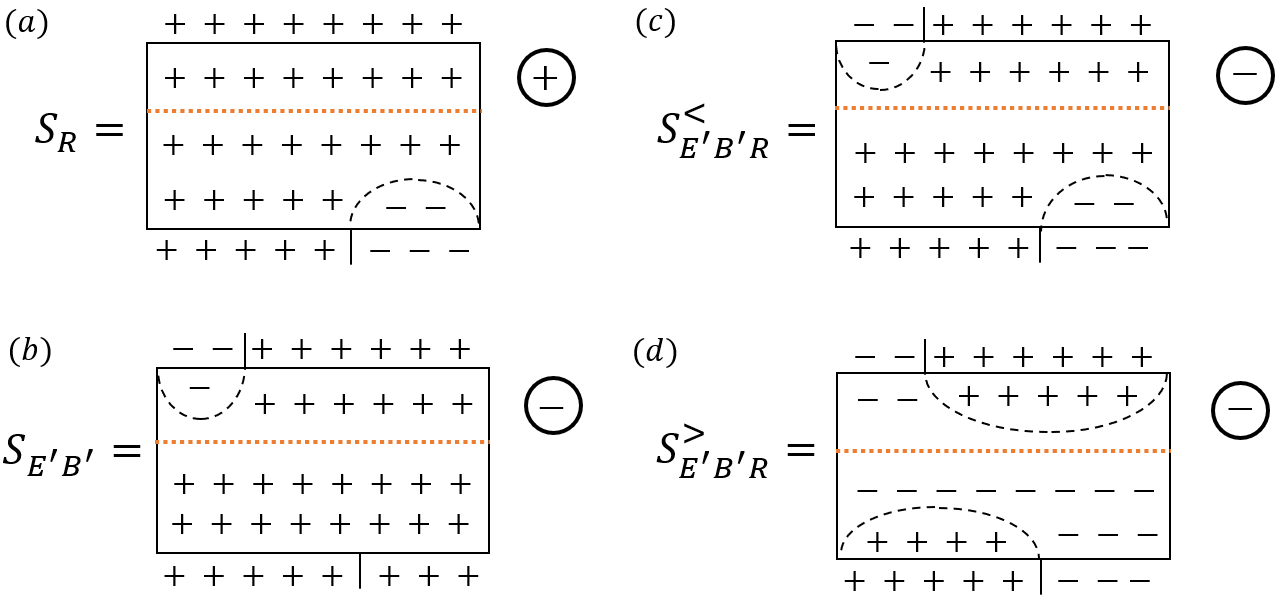}
    \caption{The domain wall configurations corresponding to various quasi entropy, (a) $S_R^{(2)}$, (b) $S_{E'B'}^{(2)}$, and (c,d) $S_{E'B'R}^{(2)}$ before and after the pinning transition. The $+$ and $-$ indicates $(\phi_1\!>\!0,\phi_2\!=\!0)$ and $(\phi_1\!=\!0,\phi_2\!>\!0)$, respectively. The orange dashed line is the boundary of the region with magnetic field, i.e., $T \!<\! t \!<\! T + T_h$. The $\pm$ sign in the circle indicates the direction of the magnetic field. Note that there is also a pinning transition in $S_{E'B'}^{(2)}$, but we only show the configuration before the transition for simplicity.}
    \label{fig:s4}
\end{figure}

Now we discuss the domain wall picture of the error threshold from the quasi-2 entropy.
As is done in the main text, we can introduce an inaccessible environment to model the record loss error.
For each measurement of the $i$-th Majorana in site $x$, if the outcome is not recorded, we can introduce an environment qutrit at $|0_E \rangle \langle 0_E |$, and the record loss is $\rho \otimes |0_E \rangle \langle 0_E | \rightarrow U_E (\rho \otimes |0_E \rangle \langle 0_E | ) U_E^\dag = \sum_{i} M_i \rho M_j^\dag \otimes |i_E \rangle \langle j_E |$, where $U_E = \sum_{i=1}^2 M_i \otimes \sum_{j=0}^2 |i_E \rangle \langle j_E |$. 
Tracing out the environment will generate an emergent magnetic field. More explicitly, the following trace,
\bea
	\Tr_E [(U_E  |0_E \rangle \langle 0_E |  U_E^\dag)^{ \otimes2}] = \sum_{i,j} M_i \otimes M_i^\dag \otimes  M_j \otimes M_j^\dag, \\
	\Tr_E [(U_E  |0_E \rangle \langle 0_E |  U_E^\dag)^{ \otimes2} \mathcal{S}] = \sum_{i,j} M_i \otimes M_j^\dag \otimes M_j \otimes M_i^\dag, 
\eea
where $\mathcal S$ is the swap operator, leads to magnetic field pointing along the direction of $\phi_1$ (the ``$+$" direction) and $\phi_2$ (the ``$-$" direction), respectively. 
In $S_{R}^{(2)}$, the environment qutrit is traced out without a swap operator, so the magnetic field is along ``$+$" direction.
On the other hand, in $S_{E'B'}^{(2)}$ and $S_{E'B'R}^{(2)}$, the environment qutrit is traced out with a swap operator, so the magnetic is along ``$-$" direction. 
This is explicitly shown in Fig.~\ref{fig:s4}.
The difference between the domain wall picture in the main text and here is the boundary condition of part $B$ in the bottom boundary. 
Because we have collected the information of $B$ represented by $C$, tracing out $C$ leads to a boundary condition along ``$+$" direction in $B$.
Including such a modification, $S_{E'B'R}^{(2)}$ reads
\bea
	S^{(2)}_{E'B'R} = - \log ( e^{-\sigma (e+ \eta) L + h T_h (1-e)  L} + e^{- \sigma (1-e) L} e^{- \sigma (1-\eta) L} ).
\eea
The above equation determines the erasure error threshold $e_c$ in the presence of the record loss,
\bea 
	h T_h (1-e_c) = 2\sigma(1-e_c-\eta).
\eea
If the measurement rate is zero, and so is the record loss error, i.e., $h=0$, we have reproduced the result of decoupling theorem.
Equivalently, if the erasing part is fixed, it is precisely the pinning transition at $h^* = \frac{2\sigma}{T_h} \frac{1-e-\eta}{1-e}$.

It is worth noting that there is also a similar pinning transition in $S_{E'B'}^{(2)}$. So including pinning transitions in both $S_{E'B'R}^{(2)}$ and $S_{E'B'}^{(2)}$, the mutual information reads
\bea
	I^{(2)}(R;E'B') =  \begin{cases}
		0 & h < h^* \\
		N[2\sigma(\eta + e -1) + h T_h (1-e)]L & h^* < h  < h^{**} \\
		2 N \sigma \eta L & h^{**} < h
	\end{cases},
\eea
where $h^{**} = \frac{2\sigma}{T_h}$ is from the pinning transition of $S_{E'B'}^{(2)}$. Thus from the domain wall picture of quasi-2 entropy, we have obtained the prediction of error threshold in the Hayden-Proskill protocol in the presence of record-loss error.

\end{document}